\documentclass[aps,prl,preprintnumbers,amsmath,amssymb,twocolumn,
epsf,superscriptaddress,groupedaddress,nofootinbib,tightenlines]{revtex4}

\usepackage{graphicx}
\usepackage{amsmath}
\usepackage{bm}

\def\beqn{\begin{eqnarray}}
\def\eeqn{\end{eqnarray}}
\def\bea{\begin{align}}
\def\eea{\end{align}}
\def\barr{\begin{array}}
\def\earr{\end{array}}
\def\btab{\begin{tabular}}
\def\etab{\end{tabular}}
\def\bite{\begin{itemize}}
\def\eite{\end{itemize}}
\def\bcen{\begin{center}}
\def\ecen{\end{center}}

\def\eq{\begin{equation}}
\def\ee{\end{equation}}

\def\q2dagger{q_2\hspace{-0.35cm}/\;}

\newcommand{\affA}{PRISMA$^{++}$ Cluster of Excellence, Institut f\"ur Kernphysik, Johannes Gutenberg-Universit\"at, Mainz, Germany}
\newcommand{\affB}{PRISMA$^{++}$ Cluster of Excellence, Institut f\"ur Physik, Johannes Gutenberg-Universit\"at, Mainz, Germany}

\begin{document}
\title{Parity violation in atoms: neutrino-mediated long range forces and finite nuclear size}
\author{Mikhail Gorchtein}\affiliation{\affA}
\author{Hubert Spiesberger
}\affiliation{\affB}
\date{\today}

\begin{abstract}
We consider neutral-current parity-violating interactions in an atom mediated by the 
exchange of a neutrino-antineutrino pair. We explicitly account for the nuclear finite 
size encoded in the nuclear form factor. Based on its general properties, we derive 
an effective neutrino-mediated potential and determine its properties at short and 
long distances. We demonstrate that, once the form factor properties are correctly 
accounted for, the range of such an effective potential corresponds to the nuclear radius, removing any sensitivity to shorter-distance contributions. 
This potential changes sign over the atom's volume, so that the correction to the effective nuclear weak 
charge induced by this interaction is tiny and does not alter the interpretation of atomic parity violation experiments.
\end{abstract}
\maketitle

Some two and a half thousand years ago Leucippus and Democritus theorized 
that matter consists of indivisible building blocks---``atoms" whose size, shape 
and the way they stick together determine the features of any material. They 
had a plan: take a very sharp knife and keep cutting a material until you get to 
its atoms that cannot be cut any more. Could they have succeeded? 
Rutherford's experiment used scattering of 5~MeV $\alpha$-particles on 
a gold foil and demonstrated that an atom has a nucleus that nearly carries the 
atom's entire mass but occupies only a tiny fraction of its volume. Could he 
have found out right away that the atomic nuclei consist of protons and 
neutrons? Hofst\"adter's experiments showed that the proton is not an 
elementary particle and has a finite size. Could he have directly proven 
that the proton is made of quarks? The answer to these naive questions 
is straightforward, yet not at all trivial: when trying to determine the properties 
of an object, those of the probe that one uses are just as important. While 
impossible to get to them with a knife, molecules and single atoms are 
routinely observed nowadays with electronic microscopes. Rutherford 
isolated protons a decade later by using a higher-energy 
$\alpha$-beam and a nucleus with a lower proton separation energy. 
With ultrarelativistic electron, proton or heavy ion beams at colliders one 
can study quarks, gluons and their interactions, whereas this is not possible 
with a 200~MeV electron beam used in Hofst\"adter's experiment. 

Recently, Ghosh et al.~\cite{Ghosh:2024ctv,Ghosh:2025ole} and later 
Flambaum and Samsonov~\cite{Flambaum:2026dhv}, considered the 
properties of an effective weak interaction potential mediated by the 
exchange of a neutrino-antineutrino pair. Because this higher-order 
contribution is UV-divergent, the authors argued that the correct 
short-distance behavior of the Standard Model has to be accounted 
for. Their results suggested that in atomic parity-violating experiments 
this hitherto neglected correction may modify the apparent value of the 
nuclear weak charge by as much as 0.8\,\%, exceeding the 0.3\,\% 
precision of the experimental measurement in cesium~\cite{Wood:1997zq}. 
The question that we raise in this letter resonates with those in the 
introduction: Can the interaction between the electron and the nucleus 
in an atom be sensitive to short-distance processes at 
$r\sim M_Z^{-1} \approx 0.002$~fm, where $M_Z \approx 91$~GeV 
is the $Z$-boson mass? The possibility of such sensitivity was dismissed back in 1930's~\cite{Tamm:1934,Iwanenko:1936} based on general arguments (for the nucleon-nucleon interaction). 
Since none of the mentioned works explicitly show how the effects of the finite nuclear size can be embedded in the potential that the nucleus exerts on the 
atomic electron, we develop a rigorous formalism to do that and apply it to atomic observables 
with and without parity violation. 
\\

\noindent
{\bf Atomic spectra with finite-size effects.} 
One of the successes of Quantum Field Theory (QFT) is the natural 
explanation of the range of an interaction via the mass of the carrier of 
this interaction. As was realized at the dawn of QFT by 
Yukawa~\cite{Yukawa:1935xg}, the exchange of a meson with 
mass $\lambda$, coupled to the interacting particles with the constant 
$g$ leads to the Yukawa potential,
\eq
V_\mathrm{Y}(r)=\frac{g^2}{4\pi}\frac{e^{-\lambda r}}{r} 
\label{eq:yukawa}
\ee
with the range $\lambda^{-1}$. 
A generic short range potential $V_\mathrm{SR}$ can be approximated 
by a $\delta$-function ($\int d^3\vec{r}\,\delta^3(\vec r)=1$),
\eq
V_\mathrm{SR} \approx \kappa_\mathrm{SR}\delta^3(\vec r), 
\quad 
\kappa_\mathrm{SR}=\int d^3\vec r\;V_\mathrm{SR}(r) .
\ee
For the Yukawa potential in Eq.~\eqref{eq:yukawa}, we find 
$\kappa_\mathrm{Y} = g^2/\lambda^2$, and the respective interaction 
range reads,
\eq
\lambda^{-1} 
= \sqrt{\frac{\kappa_\mathrm{Y}}{\left.4\pi [rV_\mathrm{Y}(r)]\right|_{r\to0}}}. 
\label{eq:range}
\ee

The constant $\kappa$ can be seen as a ``cumulative strength" 
of a short range potential over the entire space (or, in the 
approximation of a very large atomic radius, the atom's volume). 
In the rest of this article, we will use the parameter $\kappa$ 
to quantify the strength of short-range corrections to the potential 
acting on the atomic electron.

A quantum-mechanical object, the potential, can be obtained from 
a QFT scattering amplitude by means of the nonrelativistic reduction, 
neglect of the energy transfer and a three dimensional Fourier 
transform. In the dispersion formalism introduced in 
Refs.~\cite{Feinberg:1968zz,Feinberg:1989ps}, the amplitude 
$A(t)$ for elastic scattering of two particles is 
analytically continued from the scattering regime (i.e., at negative 
values of the momentum transfer $t$) to the crossed channel 
at positive momentum transfer. 
Its discontinuity $\mathrm{Disc}[A(t)] \equiv A(t+i0) - A(t-i0)$ in the crossed channel is used to obtain the potential via,
\eq
V(r) = \frac{1}{8\pi^2r} 
\int_0^\infty dt e^{-\sqrt{t}r} \mathrm{Disc}[A(t)]. 
\ee

The Coulomb potential for an electron with the charge $-e$ in the 
field of a pointlike charge $Ze$, e.g., follows 
from the Feynman diagram with one-photon exchange, 
\beqn
A_C(t) = \frac{-ie^2Z}{t+i\epsilon} 
&\to& 
V_C(r) = -\frac{Z\alpha}{r} 
\eeqn
with $\alpha=e^2/(4\pi)$ the fine structure constant.

The Coulomb potential for a point-charge leads to a degenerate 
spectrum of the hydrogen atom: the energy levels only depend on the principal quantum number $n$. The degeneracy is lifted 
by QED radiative corrections and the nuclear finite size~\cite{Eides:2007exa}. 
We introduce the finite size effect in the scattering amplitude via a 
form factor $F(t)$ and use the superscript ``FS" to indicate it,
\beqn
A^\mathrm{FS}_C(t)=\frac{-ie^2ZF(t)}{t+i\epsilon}.
\eeqn

To proceed with the dispersion calculation of the potential, we need 
to define the form factor in terms of its analytical structure. 
\begin{figure}[b]
	\centering
	\includegraphics[width=0.5\columnwidth]
    {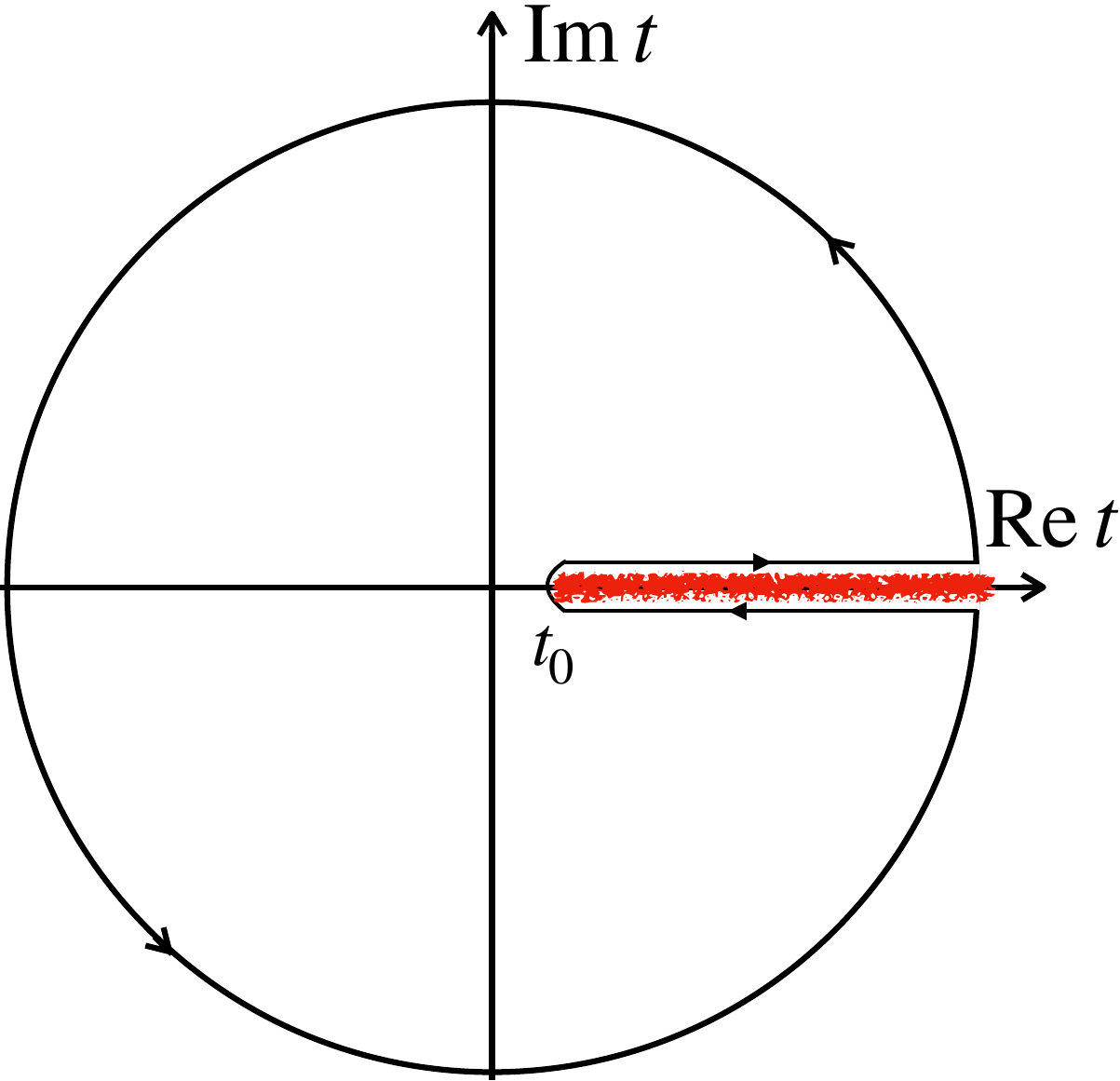}
	\caption{Analytical structure of the form factor on the complex $t$-plane.}
	\label{fig:contour}
\end{figure}
As displayed in Fig.~\ref{fig:contour}, the form factor is an 
analytical function on the entire complex $t$-plane with the exception 
of a unitarity cut along the positive $t$-axis.
The form factor obeys a dispersion representation,
\beqn
F(t) = \int_{t_0}^\infty \!\!\frac{dt'\rho(t')}{t'-t-i\epsilon} , \;\; 
\mathrm{Disc}[F(t)]=2\pi i\rho(t)\theta(t-t_0),
\eeqn
where $t_0>0$ is the threshold of the spectral function. Its exact 
value is of no relevance for our purpose. The spectral function 
obeys several general constraints:
\beqn
1.&&\int(dt/t)\rho(t)=F(0)=1\label{eq:norm} ;
\\
2.&&\int(dt/t^2)\rho(t)=F'(0)=R^2/6\label{eq:radius} ;
\\
3.&&\int dt\,t^{n-1}\rho(t)=0,\; 
\left\{\begin{array}{l}
1\leq n\leq m,\label{eq:SCR}\\
\mathrm{if}\;|F(|t|\to\infty)|< |t|^{-m} . 
\end{array} 
\right. ~~~
\eeqn
The former two constraints fix the normalization, and the slope 
in terms of the nuclear radius. The conditions in Eq.~\eqref{eq:SCR} 
are called superconvergence relations (SCR) and encode the 
asymptotic behavior of $F$. For the special case of a dipole 
form factor one has $m=1$ and there is only one SCR with $n=1$. 
Taking into account the form factor, the Coulomb potential is 
modified and reads
\beqn
&&\mathrm{Disc}\left[\frac{-ie^2ZF(t)}{t+i\epsilon}\right] 
= - 2\pi e^2Z\left[\delta(t)-\frac{\rho(t)}{t}\right] 
\nonumber\\
&&\to V^\mathrm{FS}_C(r) 
= - \frac{Z\alpha}{r}\left[1-\int_{t_0}^\infty\frac{dt}{t}\rho(t)e^{-\sqrt tr}\right]. 
\eeqn
The extra term within the brackets represents the finite-size correction 
to the Coulomb potential which we denote as $\delta V^\mathrm{FS}_C(r)$. 
It has a short range and we obtain,
\eq
\kappa[\delta V_C^\mathrm{FS}] 
= 4\pi Z\alpha\! \int\limits_0^\infty \!rdr 
\int\limits_{t_0}^\infty\!\frac{dt}{t}\rho(t)e^{-\sqrt tr} 
= 4\pi Z\alpha \frac{R^2}{6},
\label{eq:deltaVFS}
\ee
where we performed $\int dr$ first and identified the remaining integral 
with the nuclear radius via Eq.~\eqref{eq:radius}. Since 
\eq
\left.r\delta V^\mathrm{FS}_C(r)\right|_{r=0} 
= Z\alpha\int_{t_0}^\infty\frac{dt}{t}\rho(t)=Z\alpha,
\ee
where we have used Eq.~\eqref{eq:range}, we obtain the 
range of the finite-size correction to the Coulomb potential 
as $\lambda^{-1}=R/\sqrt{6}$, so that the potential takes the 
Yukawa or contact form,
\eq
\delta V_C^\mathrm{FS}(r) 
= Z\alpha\frac{e^{-\sqrt6\frac{r}{R}}}{r} 
\approx \frac{2}{3}\pi Z\alpha R^2\delta^3(\vec r).
\ee

With this effective potential, the finite size contribution to the Lamb shift, 
the energy splitting between the $2P$ and $2S$ states of a hydrogen-like 
atom, reads~\cite{Eides:2007exa,Antognini:2022xoo} 
\eq
\Delta E_{2P-2S}^\mathrm{FS}=\frac{(Z\alpha)^4 m_r^3}{12}R^2+\dots 
\ee
(we use the spectroscopy notation $n \ell$, where $n$ and $\ell$ 
stand for the principal and angular quantum numbers, respectively, 
and $S,P,D, \dots$ correspond to $\ell=0,1,2,\dots$). The reduced 
mass of an atom is given by $m_r=mM/(m+M)$ where $M$ is the 
mass of the atomic nucleus and $m$ the lepton mass, $m_e$ for 
an ordinary atom and $m_\mu$ for a muonic one. The ellipses 
indicate further subleading finite size contributions.

Higher-order QED radiative corrections also contribute to the Lamb shift. 
Beyond the perturbative expansion in powers of $Z\alpha$, they follow 
a hierarchy in terms of the range of the interaction. This hierarchy leads, 
amongst others, to a very different impact of the electronic vacuum polarization 
(eVP, in atomic physics known as the Uehling potential~\cite{Uehling:1935uj}) 
on the Lamb shift in ordinary and muonic hydrogen. Its contribution 
is proportional to the ratio of the electron's Compton wave length 
$(m_e)^{-1}$ and the size of the $2S$-orbital $(\alpha m_r/2)^{-1}$, 
to the third power~\cite{Eides:2007exa,Antognini:2022xoo}.
For hydrogen $\alpha m_r/(2m_e)\approx \alpha/2$, while for muonic 
hydrogen $(\alpha m_r/2m_e) \approx 1$. As a result, eVP contributes 
only a fraction of the Lamb shift in hydrogen, but $\approx 98\,\%$ of that 
in muonic hydrogen. This example illustrates that the range of an 
effective interaction can alter naive expectations of the perturbative 
expansion. Importantly, higher-order QED effects and finite nuclear size 
lead to short-range corrections to the point-Coulomb potential. 
The hierarchy of the range of Coulomb\,/\,Uehling\,/\,FS potentials 
for the hydrogen atom reads $(\alpha m_e)^{-1}\gg m_e^{-1}\gg R/\sqrt6$, 
while for muonic hydrogen it becomes $(\alpha m_\mu)^{-1} 
\approx m_e^{-1}\gg R/\sqrt6$. The Coulomb interaction remains 
long-range even in the presence of short-range corrections. This 
is the point that we wish to emphasize here: the range of a sum 
of various interactions is that of the longest range force. 
\\

\noindent
{\bf Parity violation in atoms with finite nuclear size.} 
In the presence of parity violation the potential acting on a left- and 
right-handed electron is slightly different, $V_C-V_Z$ and $V_C+V_Z$, 
respectively. The parity-violating potential $V_Z$ arises from the 
exchange of the $Z$-boson. Because the typical momenta in an atom 
are much smaller than the $Z$-mass, the resulting Yukawa potential 
with the range $M_Z^{-1}\approx0.002$~fm can well be approximated 
by a $\delta$-function,
\beqn
A_Z(t)
&=&
\left(\frac{g}{4c_W}\right)^2\frac{iQ_W}{t-M_Z^2+i\epsilon},
\label{eq:4FZ} 
\\
V_Z(r)
&=&
\frac{g^2Q_W}{64\pi c^2_W}\frac{e^{-M_Zr}}{r} 
\approx 
V_{4F}(r)=\frac{G_FQ_W}{2\sqrt2}\delta^3(\vec r).
\nonumber
\eeqn
The electroweak coupling $g$ is related to the Fermi constant, well
known from the muon lifetime, and to the $W$-boson mass by
$G_F = g^2 / (4\sqrt2M_W^2)$. The cosine of the weak mixing
angle is given by $c_W=\cos\theta_W=M_W/M_Z$. The nuclear
weak charge $Q_W$ of a nucleus of atomic number $Z$ and
mass number $Z+N$ is expressed as $Q_W=Z(1-4s^2_W)-N$,
and $s^2_W=\sin^2\theta_W$. From Eq.~\eqref{eq:4FZ} we can
read off the cumulative strength $\kappa_Z = G_F Q_W / (2\sqrt2)$,
i.e.\ $\kappa_Z$ measures the weak charge.

Since the typical value of the nuclear radius is a few fermi, 
we expect, following the arguments of the previous section, 
that the range of the weak interaction of an atomic electron 
with a realistic nucleus should be given by the nuclear radius, 
rather than by the inverse $Z$-boson mass. Below, we disregard 
the difference between the charge and weak form factors and 
radii for simplicity. The finite nuclear size modifies the PV scattering 
amplitude as
\beqn
A_Z^\mathrm{FS}(t)=\frac{g^2}{16c^2_W}\frac{iQ_WF(t)}{t-M_Z^2+i\epsilon} 
\approx 
A_{4F}^\mathrm{FS}(t)=\frac{-iG_FQ_WF(t)}{2\sqrt2}. 
\nonumber\\
\eeqn
The respective discontinuity and the potential read
\beqn
&&\mathrm{Disc}\left[A_{4F}^\mathrm{FS}(t)\right] 
= 2\pi\frac{G_F}{2\sqrt2}Q_W\rho(t) 
\\
&&V_{4F}^\mathrm{FS}(r)=\frac{G_F}{2\sqrt2}Q_W 
\int_{t_0}^\infty dt\rho(t)\frac{e^{-\sqrt{t}r}}{r}, 
\nonumber
\eeqn
and we obtain for the cumulative strength parameter,
\beqn
\kappa_{4F}^\mathrm{FS} 
= \frac{G_F}{2\sqrt2}Q_W\int_{t_0}^\infty \frac{dt }{t}\rho(t) 
= \frac{G_F}{2\sqrt2}Q_W=\kappa_Z,
\eeqn
where the normalization constraint on $\rho$ was used. 
We observe that the cumulative weak charge is not affected 
by the finite nuclear size. To determine the shape of the potential 
one would need to know the spectral function. However, we can 
deduce some of its properties from those of the spectral function.
The SCR $\int\rho(t)dt=0$ implies that 
$rV_{4F}^\mathrm{FS}(r)|_{r\to0}=0$, so that the 
resulting weak potential is not of the Yukawa form and remains 
finite at short distances. While the point-nucleus weak potential 
in Eq.~\eqref{eq:4FZ} is divergent at $r=0$, the finite nuclear size 
removes this divergent behavior completely. This is in accord with 
the observation that a sum of potentials with different ranges assumes 
the longest range among them.

Because the weak interaction has such a short range, radiative 
corrections to the tree-level process may have a longer range. 
A $t$-channel exchange of a $\nu\bar\nu$-pair in the 4-Fermi 
theory gives rise to the following UV-divergent amplitude evaluated 
using dimensional regularization: 
\beqn
A_{\nu\bar\nu}(t) 
= i\frac{G_F^2Q_Wt}{144\pi^2}\left[5+3\Delta_E+3\ln\frac{\mu^2}{-t}\right], 
\label{eq:nunu-ampl}
\eeqn
with $\Delta_E=1/\epsilon-\gamma_E+\ln4\pi$ and $\epsilon\to0^+$. 
Its discontinuity arises from the logarithm, Disc$\left[\ln\frac{\mu^2}{-t}\right] 
=2\pi i$, 
so that the potential is UV-finite,
\beqn
V_{\nu\bar\nu}(r) 
= - \frac{G_F^2Q_W}{192\pi^3r}\int_0^\infty dt \,te^{-\sqrt t r} 
= - \frac{G_F^2Q_W}{16\pi^3r^5}, 
\label{eq:VSE}
\eeqn
as pointed out in Refs.~\cite{Feinberg:1968zz,Feinberg:1989ps}. 
This interaction has an unphysical behavior at short distances, 
so that $\kappa_{\nu\bar\nu}=\infty$, indicating that the calculation 
missed something important. The authors of 
Refs.~\cite{Ghosh:2024ctv,Ghosh:2025ole} conclude that one 
has to open the 4-Fermi interaction and explicitly include the 
heavy weak bosons. In doing so, they find that the divergent 
behavior at short distances is indeed removed. Because it is 
the inclusion of the $Z$, $W$-boson exchange that cures the 
UV behavior, they naturally find that the correction to the weak 
charge scales as $O(G_FM_Z^2)\sim10^{-3}$, at the level of, 
or exceeding the experimental precision. Below, we show that 
this claim is based on the neglect of the nuclear finite size 
which, once correctly accounted for, relegates the relative effect 
to a tiny correction.

To account for the finite nuclear size, the amplitude in 
Eq.~\eqref{eq:nunu-ampl} is supplemented by the nuclear form factor, 
\beqn
A_{\nu\bar\nu}^\mathrm{FS}(t) 
= i\frac{G_F^2Q_WtF(t)}{144\pi^2}\left[5+3\Delta_E+3\ln\frac{\mu^2}{-t}\right],
\eeqn
and the potential reads
\beqn
V^\mathrm{FS}_{\nu\bar\nu}(r)
&=& 
- \frac{G_F^2Q_W}{192\pi^3r}\int_0^\infty dt\,te^{-\sqrt t r} 
\label{eq:VSEFS}\\
&\times& \left[\mathrm{Re}F(t)+\left(\frac{5}{3}+\Delta_E+\ln\frac{\mu^2}{t}\right)\rho(t)\right]. 
\nonumber
\eeqn

With respect to Eq.~\eqref{eq:VSE}, the factor $t$ that led to the divergent behavior 
is now tempered by the form factor, and in addition the spectral function appears. 
We compute the respective cumulative strength,
\beqn
\kappa^\mathrm{FS}_{\nu\bar\nu}
=-\frac{G_F^2Q_W}{48\pi^2} 
\int\limits_{0}^\infty dt\left[\mathrm{Re}F(t)+\ln\frac{\mu^2}{t}\rho(t)\right],
\eeqn
where the UV-divergent terms drop out due to the SCR $\int\rho(t)dt=0$. 
For the same reason, the above result is independent of the renormalization 
scale $\mu$ which is only kept to make the argument of the logarithm 
dimensionless. Note that only the first term needs to be integrated 
from $0$, while the spectral function's support starts at $t_0>0$. 
We rewrite the first term as
\beqn
&&\int_0^\infty dt\,\mathrm{Re}\,F(t)=\int_0^\infty dt\,\mathcal{P}\!\!\int_{t_0}^\infty\frac{dt'\rho(t')}{t'-t}
\nonumber\\
&&=\int_0^\infty dt\,\mathcal{P}\!\!\int_{t_0}^\infty dt'\rho(t') 
\left[\frac{1}{t'-t}+\frac{1}{\lambda^2+t}\right]
\nonumber\\
&&=\int_{t_0}^\infty dt'\rho(t')\ln(t'/\lambda^2),
\eeqn
where in the second line we added a constant term in $t'$ with 
an arbitrary mass $\lambda$ in the square bracket. Due to the 
SCR $\int\rho(t)dt=0$, this term integrates to $0$ and has no 
effect on the final result but renders the integral over $t$ finite. 
Because the spectral function vanishes sufficiently fast for large 
$t$, we can safely change the order of integrations above. 
The potential in Eq.~\eqref{eq:VSEFS} is not identically zero,  but it integrates to zero over the entire space,
\beqn
\kappa^\mathrm{FS}_{\nu\bar\nu} 
= -\frac{G_F^2Q_W}{48\pi^2}\int_{t_0}^\infty dt\ln({\mu^2}/{\lambda^2})\rho(t)=0.
\eeqn

The cancellation goes even beyond the expectation 
$\kappa^\mathrm{FS}_{\nu\bar\nu}\propto G_FQ_W\times G_F/R^2$ 
and relies on two assumptions: 
\eq
1.\,\int\limits_{t_0}^\infty dt\,\rho(t)={0}, 
\quad\mathrm{and}\quad 
2.\; \kappa=4\pi\int\limits_0^\mathbf{\infty} r^2dr\,V(r).
\ee

The first assumption follows from the asymptotic behavior 
of the weak form factor at large $t$. The pQCD scaling predicts 
that the nucleon form factor should drop at least as 
$t^{-2}$~\cite{Lepage:1979za}, and the meson form factor 
as $\alpha_s(t)/t\sim (t\ln t)^{-1}$, with $\alpha_s$ the running 
strong coupling constant~\cite{Lepage:1979zb}. Nuclear form 
factors drop much faster. With this, the use of the SCR is justified 
for any hadron. 

Assumption 2 means that the interaction strength $\kappa$ is 
determined by integrating the potential over the entire space, 
rather than the atomic size, $r\leq (Z\alpha m)^{-1}$. 
We correct this by defining
\eq
\tilde\kappa^\mathrm{FS}_{\nu\bar\nu} 
= \int_\mathrm{atom\,volume}d^3\vec r \,V^\mathrm{FS}_{\nu\bar\nu}(r),
\ee
and note that
\eq
\int_0^{(Z\alpha m)^{-1}} r\,dr\,e^{-\sqrt tr} 
= 
\frac{1}{t}\left[1-\frac{\sqrt t}{Z\alpha m}e^{-\frac{\sqrt t}{Z\alpha m}}\right].
\ee

The constant term gives a zero contribution to 
$\kappa^\mathrm{FS}_{\nu\bar\nu}$, as shown before. 
Since $Z\alpha m R\ll1$ (e.g., for ${}^{133}$Cs, $R\approx4.8\,$~fm 
and $Z=55$, hence $Z\alpha m R\approx0.005$), we can safely 
set $F(t)=1$ and $\rho(t)=0$, and obtain 
\beqn
\tilde\kappa^\mathrm{FS}_{\nu\bar\nu} 
= \frac{G_F^2Q_W}{48\pi^2}\int_{0}^\infty \!\!\!\! 
dt\frac{\sqrt te^{-\frac{\sqrt t}{Z\alpha m}}}{Z\alpha m} 
= \frac{G_F^2Q_W}{12\pi^2}(Z\alpha m)^2 .
\eeqn

Since the parameters $\kappa$ and $\tilde\kappa$ are related to the 
correction to the nuclear weak charge, we rewrite
\beqn
\frac{\delta Q_W^{Z,N}}{Q_W^{Z,N}} 
= \frac{\tilde\kappa^\mathrm{FS}_{\nu\bar\nu}}{\kappa_Z} 
= \frac{G_F(Z\alpha m)^2}{3\sqrt2\pi^2}\sim10^{-14}\left(\frac{Z}{55}\right)^2.
\eeqn

We find that the sensitivity of atomic PV observables to order $G_F^2$ 
in the electroweak interaction, mediated by the exchange of two neutrinos, 
is limited by the atomic size $r_a$, and the relative impact on the 
effective nuclear weak charge scales as 
$\kappa^\mathrm{FS}_{\nu\bar\nu}/\kappa_Z\sim G_F/r_a^2$. 
This conclusion relies on introducing the nuclear finite size effects 
keeping an explicit account of general theory constraints, such as 
analyticity of the form factor and the pQCD scaling rules. 
All contributions to the potential from Feynman diagrams with a 
$\nu\bar\nu$ pair, originating from vertex corrections at the electron 
or quark side considered in Refs.~\cite{Ghosh:2024ctv,Ghosh:2025ole} 
follow the same constraint because in the 4-Fermi theory (i.e., in the 
limit $|t|\ll M_{Z,W}^2$) the contributions of all relevant diagrams are 
$\propto G_F^2t$. Besides, the long-range nature of the induced 
$\nu\bar\nu$-mediated potential is a misconception even in the 
calculation of Refs.~\cite{Ghosh:2024ctv,Ghosh:2025ole,Flambaum:2026dhv}: 
the entire effect comes from $t\sim M_Z^2$, in contrast to the 
physical picture put forward by the authors. The nuclear form factor 
cuts off this range of $t$, removing the claimed effect almost entirely. 
Interestingly, once analyticity and the asymptotic behavior of the 
form factor are accounted for, even the sensitivity to $t\sim R^{-2}$ 
is lost, and the only scale that matters is $t\sim r_a^{-2}$ which leaves 
just a tiny effect of the order $G_F/r_a^2$. Thus, as pointed out in 
the introduction, the resolution of the probe defines the limits of the 
precision one can reach. The coarse resolution of an atomic probe, 
$\sim r_a$, prevents one from resolving the effective higher-order 
electroweak interactions at shorter distances. 
\\

A final note is in order. While Ref.~\cite{Flambaum:2026dhv} only 
considers ordinary atoms, Refs.~\cite{Ghosh:2024ctv,Ghosh:2025ole} 
include exotic, purely leptonic systems such as positronium and 
muonium. Because leptons have no intrinsic size, the conclusions 
of the cited works are likely to remain intact for these systems. 
From the practical point of view, however, a measurement of parity 
violation in positronium or muonium remains a distant possibility. 
\\

\noindent
{\bf Acknowledgments.} 
The authors are grateful to J.~Erler, V.~Flambaum and V.~Pascalutsa for discussions, and to A.~V.~Borisov for pointing us to Refs.~\cite{Tamm:1934,Iwanenko:1936}.
We acknowledge support by the 
Cluster of Excellence ``Precision Physics, Fundamental Interactions, 
and Structure of Matter" (PRISMA$^{++}$ EXC 2118/2) funded by the 
German Research Foundation (DFG) within the German Excellence 
Strategy (Project ID 390831469). 
M.G.\ also acknowledges support by the Deutsche Forschungsgemeinschaft 
(DFG) - GO 2604/3-2, Projektnummer 495329596. 
\\


\begin{thebibliography}{99}

\bibitem{Ghosh:2024ctv}
M.~Ghosh, Y.~Grossman, C.~Sieng and B.~Yu,
Phys. Rev. D \textbf{112} (2025) no.11, 113004
doi:10.1103/PhysRevD.112.113004
[arXiv:2410.19059 [hep-ph]].

\bibitem{Ghosh:2025ole}
M.~Ghosh, Y.~Grossman, C.~Sieng and B.~Yu,
Phys. Rev. Lett. \textbf{135} (2025) no.25, 251803
doi:10.1103/g7tx-srnn
[arXiv:2512.07938 [hep-ph]].

\bibitem{Flambaum:2026dhv}
V.~V.~Flambaum and I.~B.~Samsonov,
[arXiv:2602.22466 [hep-ph]].

\bibitem{Wood:1997zq}
C.~S.~Wood, S.~C.~Bennett, D.~Cho, B.~P.~Masterson, J.~L.~Roberts, C.~E.~Tanner and C.~E.~Wieman,
Science \textbf{275} (1997), 1759-1763
doi:10.1126/science.275.5307.1759

\bibitem{Tamm:1934}
I.~Tamm, 
Nature 134, 1010-1011 (1934)
https://doi.org/10.1038/1341010c0

\bibitem{Iwanenko:1936}
D.~Iwanenko und A.~Sokolow, 
Z. Physik 102, 119-131 (1936)
[https://doi.org/10.1007/BF01336835]

H.~Yukawa,
Proc. Phys. Math. Soc. Jap. \textbf{17} (1935), 48-57
doi:10.1143/PTPS.1.1

\bibitem{Yukawa:1935xg}
H.~Yukawa,
Proc. Phys. Math. Soc. Jap. \textbf{17} (1935), 48-57
doi:10.1143/PTPS.1.1

\bibitem{Feinberg:1968zz}
G.~Feinberg and J.~Sucher,
Phys. Rev. \textbf{166} (1968), 1638-1644
doi:10.1103/PhysRev.166.1638

\bibitem{Feinberg:1989ps}
G.~Feinberg, J.~Sucher and C.~K.~Au,
Phys. Rept. \textbf{180} (1989), 83
doi:10.1016/0370-1573(89)90111-7

\bibitem{Eides:2007exa}
M.~I.~Eides, H.~Grotch and V.~A.~Shelyuto,
Springer Tracts Mod. Phys. \textbf{222} (2007), pp. 1-262
Springer-Verlag, 2007,
ISBN 978-3-540-45269-0, 978-3-540-45270-6
doi:10.1007/3-540-45270-2

\bibitem{Antognini:2022xoo}
A.~Antognini, F.~Hagelstein and V.~Pascalutsa,
Ann. Rev. Nucl. Part. Sci. \textbf{72} (2022), 389
doi:10.1146/annurev-nucl-101920-024709
[arXiv:2205.10076 [nucl-th]].

\bibitem{Uehling:1935uj}
E.~A.~Uehling,
Phys. Rev. \textbf{48} (1935), 55-63
doi:10.1103/PhysRev.48.55

\bibitem{Lepage:1979za}
G.~P.~Lepage and S.~J.~Brodsky,
Phys. Rev. Lett. \textbf{43} (1979) no.21, 545-549
[erratum: Phys. Rev. Lett. \textbf{43} (1979), 1625-1626]
doi:10.1103/PhysRevLett.43.1625.2

\bibitem{Lepage:1979zb}
G.~P.~Lepage and S.~J.~Brodsky,
Phys. Lett. B \textbf{87} (1979), 359-365
doi:10.1016/0370-2693(79)90554-9

\end{thebibliography}
\end{document}